\documentclass[twoside,fleqn]{article}
\usepackage{espcrc2,psfig}

\newcommand{\be}{\begin{equation}}
\newcommand{\ee}{\end{equation}}

\newcommand{\AmS}{{\protect\the\textfont2
  A\kern-.1667em\lower.5ex\hbox{M}\kern-.125emS}}

\title{Ultra High Energy Cosmic Rays}

\author{V. Berezinsky\thanks{Invited talk at TAUP-97}
\\INFN, Laboratori Nazionali del Gran Sasso,
             I--67010 Assergi (AQ), Italy and Institute for Nuclear 
Research, Moscow, Russia}       
\begin{document}

\begin{abstract}
The current status of Ultra High Energy Cosmic Rays (UHECR) is reviewed, with 
emphasis given to theoretical interpretation of the observed events. 
The  galactic and extragalactic origin, in case of astrophysical 
sources of UHE particles, have the problems either with 
acceleration to the observed energies or with the fluxes and spectra. 
Topological defects can naturally produce particles with energies as 
observed and much higher, but in most cases fail to produce the observed 
fluxes. Cosmic necklaces and monopole-antimonopole pairs are identified as 
most plausible sources, which can provide the observed flux and spectrum.
The relic superheavy particles are shown to be clustering in the Galactic 
halo, producing UHECR without Greisen-Zatsepin-Kuzmin cutoff. The Lightest 
Supersymmetric Particles are discussed as UHE carriers in the Universe.   

\end{abstract}

\maketitle

\section{\bf Introduction }

Cosmic rays (CR) are observed in a wide energy range, starting from 
subGeV energies and up to $3\cdot 10^{20}~eV$ (see Fig.1). Apart from 
the highest energies, these 
particles are accelerated in our Galaxy, most probably, by shocks 
produced by SN II explosions. Up to energy $10^{15} - 10^{16}~eV$ 
the CR flux is dominated by protons, at higher energies 
CR have the mixed composition, and there are indications that at energies 
about $\sim 10^{17}~eV$ iron nuclei dominate in the CR flux. In a wide 
range of energies from $1~GeV$ up the $3\cdot 10^{15}~eV$ the spectrum
is power-law $ \sim E^{-2.65}$, at energy $ 3\cdot 10^{15}~eV$ the 
spectrum steepens and becomes $\sim E^{-3.1}$ at $E > 10^{17}~eV$. 
At $E \geq 10^{19}~eV$ a new more flat component appears (see Fig.1). The 
highest energies detected so far are $2-3\cdot 10^{20}~eV$. 

The first steepening of the spectrum (the knee) at energy 
$3\cdot 10^{15}~eV - 1\cdot 10^{16}~eV$ is usually explained by 
inefficient confinement of CR in the Galaxy. This process must be 
accompanied by enrichment of heavy nuclei in CR composition at 
energy $\sim 10^{15}~eV$ and above. There are some indications that such 
enrichment is really observed.

There is no universal definition for Ultra High Energy Cosmic Rays (UHECR). 
Sometimes 
this term is applied for $E > 1\cdot 10^{17}~eV$ or $E> 1\cdot 10^{18}~eV$.
I shall use this term for $E > 1\cdot 10^{19}~eV$, where 
the new flat component appears.

It is natural to think that this component has extragalactic origin, 
though, in principle, very large halo with regular magnetic field can confine 
particles of these energies, especially if they are heavy nuclei.
UHECR of extragalactic origin have a signature called the 
Greisen-Zatsepin-Kuzmin (GZK) cutoff \cite{GZK}. This 
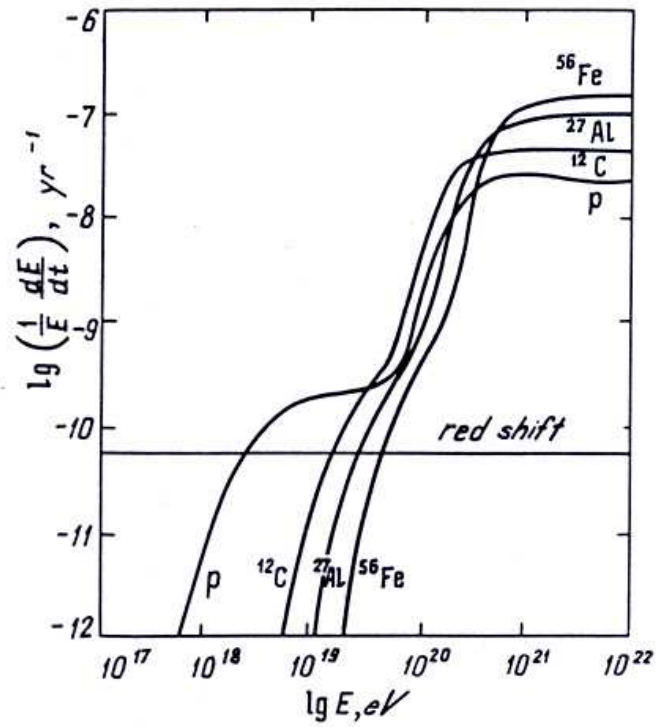
phenomenon is caused by energy losses of UHE protons due to pion production
in collisions with microwave photons. The energy losses start sharply 
increasing at $E \sim 3\cdot 10^{19}~eV$ (Fig.2). This energy is 
connected with energy of the spectrum steepening ("cutoff") in the 
model-dependent way. 
In case the sources are distributed uniformly in the Universe (standard 
assumption), the steepening starts at $E_{bb} \approx 3\cdot 10^{19}~eV$.
The flux at $E>E_{bb}$ is produced by nearby sources. If there is a 
local enhancement of the sources, $E_{bb}$ increases \cite{BBDGP};in case  
the sources are located at large distances, $E_{bb}$ decreases and steepening 
is exponential. It is more convenient to characterize  steepening by energy  
$E_{1/2}$ \cite{BBDGP}, where the flux becomes half of the power-law 
extrapolation of unmodified flux. In case of uniform distribution of 
the sources $E_{1/2} \approx 5.8\cdot 10^{19}~eV$ \cite{BBDGP} for a wide 
range of exponents $\gamma$ of generation spectrum.

Apart from GZK cutoff, there may be two more signatures  of extragalactic 
cosmic rays: a bump and a dip in differential spectrum which precede the 
cutoff \cite{HS85,BG87}. The bump is a consequence of a number conservation 
of protons in the spectrum: protons loose energy and are accumulated 
before the cutoff. The dip is formed due to pair-production ($e^+e^-$)  
energy losses of UHE proton. The both features  
show up most clearly in the differential spectrum of a single distant 
source in the case of a flat generation spectrum. In diffuse spectra (from 
many sources) these features are weak or absent. 



\noindent
{\em UHE nuclei} spectra exhibit steepening ("cutoff") approximately at 
the same energy 
as protons, though due to different physical processes (see \cite{BBDGP}
for a review). The relevant energy losses are caused by photodisintegration 
of nuclei at collisions with microwave photons, and the steepening energy
is determined by energy, when photodisintegration energy-losses start 
to dominate over adiabatic ones (Fig.2.).

{\em UHE photons} with $E_{\gamma} \sim 10^{19} - 10^{22}~eV$ have 
an absorption length less than $10~Mpc$, mainly due to interaction with 
radio-background \cite{B70,PB}.

The observation of cosmic ray particles with energies higher than 
$10^{20}~eV$ gives a serious challenge to our understanding of origin 
of UHECR: What are the mechanisms of acceleration? Why the GZK cutoff 
is absent?
\section{\bf Observational data}
The compilation \cite{akeno} of observational data for UHECR is given 
in Fig.3. Two highest energy events correspond to energies 
$3\cdot 10^{20}~eV$ (the Fly's Eye event \cite{HFE}) and $2\cdot10^{20}~eV$ 
(the AGASA event \cite{HAGASA}). 
These energies are well above the GZK cutoff for 
uniformly distributed extragalactic sources. The particles with energies 
above $10^{20}~eV$ were observed in the past at the Haverah Park array 
\cite{HP} and Sydney array \cite{Syd}. The latter detector has 
observed eight showers with $E>1\cdot 10^{20}~eV$. It operated from 
1968 up to 1979 and had large area, $87~km^2$. The scintillator 
detectors were at 2m underground and hence only muon component of 
showers was measured. It is interesting to re-analyze the data 
using the new simulations for muon distribution \cite{Lins97}.

\noindent
One shower with $E\approx 1.2\cdot 10^{20}~eV$ was observed at the Yakutsk 
array 
\cite{Gl,Yak}, though eight showers were expected if the Haverah Park data are 
correct. 

{\em Anisotropy} of UHECR is not reliably observed. Some analyses 
(e.g. \cite{HP} and \cite{SW}) indicate the excess of particles from 
Local Supercluster (LS) plane, while observations by AGASA 
\cite{AGASA1} are consistent with a uniform distribution. On other 
hand AGASA has observed\cite{AGASA1} clustering of UHE events: three pairs 
of showers with angular separation less than $2.5^{\circ}$ (for analysis 
see \cite{SLO}).
 
{\em Chemical composition} as found from analysis of the Fly's Eye data
\cite{FE} is characterized by a change from predominantly heavy nuclei 
(iron) to the light nuclei at $E \sim 3\cdot 10^{17}~eV$. The fraction of 
protons increases with energy and reaches $90\%$ at $10^{19}~eV$ \cite{FE}.  
The change of the chemical composition at $E \sim 3\cdot 10^{17}~eV$ 
was not found in the muon data in AGASA experiment \cite{AGASA2}.
The data of Yakutsk array \cite{Yak1} also favors the proton composition 
at the highest energies.

\section{\bf Galactic origin of UHECR}

There are two difficulties in an attempt to explain the observed events 
by sources in the Galaxy: the maximum acceleration energy is considerably 
less than $3\cdot 10^{20}~eV$ and galactic magnetic fields are too 
weak to isotropize UHE particles.

{\em Acceleration} to very high energies  in the Galaxy can occur at the SN 
shocks, at galactic wind terminal shock and in young pulsars. 

The maximum energy for acceleration by shocks in the interstellar medium 
does not exceed $10^{16}~eV$ \cite{Ces}. It can be higher when the SN shock 
propagates through the region of presupernova stellar wind with strong 
magnetic field \cite{BiVo}, though the maximum energy is still less than 
$3\cdot 10^{18}~eV$ \cite{BC}. The galactic wind is expected to be 
terminated by a 
standing shock, where, in case of extreme values of 
parameters used in ref.(\cite{JM}), particles can be accelerated up to energies 
$E \sim 10^{20}~eV$ for iron nuclei. Most probably this energy is an order 
of magnitude less (see \cite{BBDGP} and references therein). 

Another potential source of acceleration is young pulsar; in this case the 
maximum energy can in principle reach $10^{19}~eV$ \cite{BBDGP}. However,
in the concrete models of pulsar magnetosphere the maximum energy is less. 

{\em Propagation} of UHECR in galactic magnetic fields was studied 
numerically in many works \cite{SC,BGR,GW,PP}. The crucial element of this 
analysis 
is presence of regular magnetic field in galactic halo. 
All workers agree 
that at the highest observed energies there must 
be very strong disc anisotropy, which obviously contradicts observations. 

We shall describe here the results of calculations of ref.(\cite{BGR}.

Magnetic field is taken according to the  model of ref.(\cite{SS}). 
Several 
versions of the halo field  is considered. The size of the halo is varied, but 
generically the large size of the halo from 10 to 30 kpc is used.
The sources of UHECR are assumed to be distributed uniformly in the disc. 
The trajectories of the antinuclei with rigidity $R=E/Z$, where $Z$ is 
a charge of a nucleus, were followed step by step in the magnetic fields 
of the disc and halo. The flux in a given direction is 
proportional to the length of trajectory {\em in the disc}. An example of 
trajectories with rigidity $3\cdot10^{18}~V$ is shown in Fig.4. The particles
are emitted from the Earth in the different directions in Galactic plane 
$z=0$, where $z$ is the height over 
galactic plane. One can see that at energy $E\sim 3\cdot 10^{18}~eV$
 protons propagate almost rectilinearly in the disc, producing thus strong 
disc anisotropy. In Fig.5 the lifetimes of the 
particles 
in the disc, $T_d$, and calculated anisotropy, $A$, are shown as a 
function of rigidity $E/Z$ for two sizes of the halo. Since  
the flux is proportional to $T_d$, one concludes from Fig.5 that at 
$E>3\cdot 10^{18}Z~eV$ the particles typically do not return from the halo 
to disc.
This conclusion is confirmed by the lower part of Fig.5, which shows 
large (disc) anisotropy  at energies $E> 3\cdot 10^{18}Z~eV$. Note, that 
these calculations are performed for extreme case of very large 
magnetic halo.

{\em Relativistic dust grains} of galactic origin can be the carriers of 
UHE signal \cite{Hay}. The Lorentz factor of a dust grain should be large 
enough $\Gamma > 10^3$ to produce a nuclear-electromagnetic shower with 
muon component as observed in e.g. Sydney or AGASA arrays. Approaching the 
sun such a grain accumulates the electric charge due to photoeffect produced 
by solar optical radiation (in the rest frame of a dust grain this is 
X-ray radiation). As a result a grain breaks up due to electric repulsion
\cite{BP1}. The dust grain hypothesis is also in contradiction with observed 
properties of UHE showers (J.Linsley).

\section{\bf Extragalactic acceleration sources}

Shock acceleration (including 
ultra-relativistic shocks) and unipolar induction are the "standard" 
acceleration mechanisms to UHE, considered in the literature.
These mechanisms can operate in the various astrophysical objects, such as
Active Galactic Nuclei (AGN), large scale structures (e.g. the shocks in
AGN jets or shocks in the clusters of galaxies), in gamma-ray bursters 
(ultra-relativistic shocks), in the accretion discs around 
massive black holes (due to large electric potentials produced by unipolar 
induction) etc. A comprehensive list of possible sources was recently 
thoroughly studied in ref.(\cite{NMA}) with a conclusion, that maximum energy 
of acceleration does not exceed $10^{19} - 10^{20}~eV$. The most promising 
source from this list is a hot spot in radiogalaxy produced by a jet 
\cite{BiSt,IA,RB}.   

A powerful jet ejected from the AGN supplies energy to a gigantic radiolobe.
A hot spot observed at the termination of the jet is interpreted as a 
location of a standing shock. This is an ideal place for acceleration of 
protons to very high energies: magnetic field is strong and the energy 
density of radiation, responsible for proton energy losses, is relatively 
small. The maximum energy can be estimated as \cite{BiSt}
\be
E_{max} = 1\cdot 10^{20}H_{-4}(R/1~kpc)v_j~eV,
\label{eq:spot}
\ee
where $H_{-4}$ is the magnetic field in units of $10^{-4}~G$, $R$ is radius 
of the shock and $v_j$ is velocity of the jet in units of sound speed.
However, the powerful radio sources are at large distances from our Galaxy,
and the maximum energy is strongly attenuated. The discussed 
sources can provide the observed flux up to energies $6-7\cdot 10^{19}~eV$.

{\em Unipolar induction} produces very large potential drop in the 
accretion discs around massive black holes (see ref.(\cite{Bla,Lov} and 
also \cite{BBDGP} for a discussion). The electrical potential 
for a rotating disc, at the distance r from a black hole is
\be 
\phi(r)=\frac{1}{\sqrt{6}}H_cr_c\ln R/r,
\label{eq:pot}
\ee
where $r_c=9\cdot 10^5(M_h/M_{\odot})$ is the radius of the last stable 
orbit for a black hole with mass $M_h$, $H_c$ is magnetic field at the last 
stable orbit, and $R$ is a radius of accretion disc. The maximum potential 
given by Eq.(\ref{eq:pot}) is $ \phi_c \sim 3\cdot10^{21}~V$ for 
$M_h \sim 1\cdot 10^9 M_{\odot}$ and $H_c \sim 1\cdot 10^4~G$.  
This mechanism is attractive because it can operate not only in AGN, but 
also for the old black holes, which lost their activity ($\dot{M}$ in the 
disc is small). Such sources can be located nearby, e.g. in the Local 
Supercluster, from where UHE protons can reach us without appreciable 
energy losses.

{\em Relativistic shocks} can in principle provide very high maximum energy
. 
A particle reflecting from a relativistically moving mirror increases its 
energy by a factor proportional to $\Gamma^2$, where $\Gamma$ is a Lorentz 
factor of the mirror. This acceleration phenomenon is known from the time 
of the pioneering work by E.Fermi \cite{Fer}. The less known phenomenon is 
capturing 
of accelerated particles behind relativistically moving shock front.
One can easily reconstruct this principle considering head-on collision of 
a particle with a transverse relativistic shock (magnetic field is 
perpendicular to the shock normal). Let us assume that 
magnetic field behind the shock is homogeneous on the scale of the particle 
Larmor radius. Then at the moment a particle finishes semi-circle, the shock 
front run away to a distance $cT_L/2$ from a particle, where $T_L$ is 
the Larmor period. The confinement described above can be illustrated by 
numerical simulation in ref.(\cite{BeKi}) shown in Fig.6. 

\noindent
In Fig.6a 
the shock velocity is non-relativistic ($0.1c$) and a particle reflects 
many times from a shock. In Fig.6b the shock is relativistic, $0.87c$, 
 and a particle is captured behind the shock.   
The same phenomenon is clearly seen in the relativistic transverse shocks 
in electron-positron plasma, ( Fig.4 from \cite {Ho}).
Somewhat more complicated, but similar mechanism 
operates for parallel shock, i.e. when magnetic field is parallel to 
the shock velocity.  

The capturing mechanism described above, does not exclude completely 
the $\Gamma^2$-regime of acceleration at relativistic shocks; it 
restricts the incident angles at which particles escape and thus 
the flux of accelerated particles. 
The most interesting objects
where  $\Gamma^2$-mechanism might operate are gamma-ray bursters
\cite{Vi,Wa}. The Lorentz factor of the shock here can reach $10^2 - 10^3$.
However, the capturing properly taken into account might dramatically 
decrease the output of accelerated particles. As to the explanation of the 
observed UHECR, the protons from cosmologically remote gamma-ray bursts 
strongly degrade in energy on the way to our Galaxy.

{\em Astrophysical sources} of observed UHECR must satisfy the 
observational constraints. The absence of GZK cutoff at 
$E_{1/2} \approx 6\cdot 10^{19}~eV$ contradicts the uniform distribution 
of the sources in extragalactic space. If the sources are located as a 
dense group around our Galaxy, $E_{1/2}$ increases with increasing of  
density contrast, i.e. the ratio of number density of sources inside the 
group and outside it \cite{BBDGP}. The Local Supercluster LS) can realize this 
possibility. This model was developed in ref.(\cite{BG}). The UHECR sources 
in LS can be the old massive black holes with large electric potential 
induced in the accretion discs by unipolar induction.  For increasing 
$E_{1/2}$ up to $1\cdot 10^{20}~eV$ the density contrast larger than $10$ 
is needed. The anisotropy can be smaller than the observed one if 
magnetic field in 
superclusters is as large as $H\sim 10^{-7}~G$ \cite{KK}.

\section{Topological defects and relic particles.}

{\em Topological defects, TD,} (for a review see \cite{Book}) can naturally 
produce particles of ultrahigh energies (UHE). The pioneering observation 
of this possibility was made by Hill, Schramm and Walker \cite{HS} (for 
a general analysis of TD as UHE CR sources see \cite {BHSS} and for a 
recent review \cite{Sigl}).

In many cases TD become unstable and decompose to constituent fields, 
superheavy gauge and Higgs bosons (X-particles), which then decay 
producing UHECR. It could happen, for example, when two segments of 
ordinary string, or monopole and antimonopole touch each other, when 
electrical current in superconducting string reaches the critical value
and in some other cases.

In most cases the problem with UHECR from TD 
is not the maximal energy, but the fluxes. One very general reason 
for the low fluxes consists in the large distance between TD. A dimension
scale for this distance is the Hubble distance $H_0^{-1}$. However, in some 
rather exceptional cases this dimensional scale is multiplied to a small 
dimensionless value $r$. If a distance between TD is larger than 
 UHE proton attenuation length, then 
the flux at UHE is typically exponential suppressed.

{\em Ordinary cosmic strings} can produce particles when a loop annihilate 
into double line \cite{BR}. The produced UHE CR flux is strongly reduced due 
to the fact that a loop oscillates, and in the process of a collapse the 
two incoming parts of a loop touch each other in one point producing thus 
the smaller loops, instead of two-line annihilation. However, this idea was 
recently revived due to recent work \cite{Vincent}. It is argued there that 
the energy loss of the long strings is dominated by production of very 
small loops with the size down to the width of a string, which immediately 
annihilate into superheavy particles. A problem with this scenario is 
too large distance between strings (of order of the Hubble distance). 
For a distance between an observer and a string being the same, the 
observed spectrum of UHE CR has an exponential cutoff at energy 
$E \sim 3\cdot 10^{19}~eV$.

Superheavy particles can be also produced when two segments of
string come 
into close contact, as in {\it cusp} events \cite{Bran}.  
This process
was studied later in ref.(\cite{GK}) with a conclusion that
the resulting cosmic ray flux is far too small.  
An interesting possibility suggested in ref.(\cite{Bran}) is the
{\em cusp} ``evaporation'' on cosmic strings. 
When the distance between two segments of the cusp
becomes of the order of the string width, the cusp may``annihilate" 
turning into high energy particles.
, which are boosted by a very large Lorentz
factor of the cusp \cite{Bran}.  
However, the resulting UHE CR flux is considerably smaller than one 
observed \cite{BBM}.  

{\em Superconducting strings} \cite{Witten} appear to
be much better suited for particle production.
Moving through cosmic magnetic fields, such strings develop
electric currents and copiously produce charged heavy particles when the
current reaches certain critical value. 
The CR flux produced by
superconducting strings is affected by some model-dependent string
parameters and by the history and spatial distribution of cosmic
magnetic fields.  
Models considered so far failed to account
for the observed flux \cite{SJSB}.

{\em Monopole-antimonopole pairs } ($M{\bar M}$) 
can form bound states and eventually
annihilate into UHE particles  \cite{Hill}, \cite{BS}.  
For an appropriate choice of the
monopole density $n_M$, this model is consistent with observations;
however, the required (low) value of $n_M$ implies fine-tuning. 
In the first phase transition $G \to H \times U(1)$ in the early 
Universe the monopoles are produced with too high density. It must then be 
diluted by inflation to very low density, precisely tuned to 
the observed UHE CR flux. 

{\em Monopole-string network} can be formed in the early Universe in the 
sequence of symmetry breaking 
\be
G \to H \times U(1) \to H \times Z_N.
\label{eq:Z-N}
\ee
For $N \geq 3$ an infinite network of monopoles connected by strings is 
formed. The magnetic fluxes of monopoles in the network are channeled into 
into the strings that connect them. The monopoles typically have additional 
unconfined magnetic and chromo-magnetic charges. When strings shrink the 
monopoles are pulled by them and are accelerated. The 
accelerated monopoles produce extremely high energy gluons, which then 
fragment into UHE hadrons \cite{BMV}. The produced flux is too small 
to explain UHE CR observation \cite{BBV}.

{\em Cosmic necklaces} are TD which are formed  
in a sequence of symmetry breaking given by Eq.(\ref{eq:Z-N}) when $N=2$. 
 The first phase transition
produces monopoles, and at the second phase transition each monopole
gets attached to two strings, with its magnetic flux channeled along the 
strings.  The resulting necklaces resemble ``ordinary'' cosmic strings
with monopoles playing the role of beads. Necklaces can evolve in such way
that a distance between monopoles diminishes and in the end all monopoles 
annihilate with the neighboring antimonopoles \cite{BV}. 

An important quantity for the necklace evolution is is the dimensionless 
ratio $r=m/\mu d$, where $m$ is the monopole mass, $\mu$ is is the string 
tension, determined by $U(1)$ symmetry scaling scale $\eta_s$, and $d$ is 
a distance between two neighbouring monopoles. In ref.(\cite{BV}) it is 
argued that in the process of the necklace evolution $r(t)$ is driven 
towards large values $r \gg 1$. The characteristic length scale, equal to
the typical separation of necklaces,  
\be
\xi \sim r^{-1/2}t, 
\label{eq:xi}
\ee
is much smaller than horizon $t$, when $r$ becomes large.

A requirement for all models explaining the observed UHE events is that 
the distance between sources must be smaller than the attenuation
length for UHE particles. Otherwise the flux at the 
corresponding energy would be 
exponentially suppressed. This imposes a severe constraint on the
possible sources.  For example, in the case of protons with energy 
$E \sim (2- 3)\cdot 10^{20}~eV$, 
the proton 
attenuation length is $19~Mpc$ . If protons 
propagate rectilinearly, there should be several sources inside this
radius;
otherwise all particles would arrive from the same direction.
If particles are strongly deflected in extragalactic magnetic fields, 
the distance to the source should be even smaller.  Therefore, the 
sources of 
the observed events at the highest energy must be at a distance 
$R\leq 15~Mpc$ in the case or protons. 

For the necklaces the distance between 
sources, given by Eq.(\ref{eq:xi}), satisfies this condition for 
$r>3\cdot 10^{4}$.
This is in contrast to other potential sources,
including supeconducting cosmic strings and powerful 
astronomical sources such as AGN, for which this condition imposes
severe restrictions.  

The diffuse fluxes of UHE protons and photons from necklaces are given in
Fig.7.
One can see that at $E \sim 1\cdot 10^{20}~eV$ the 
gamma-ray flux is considerably lower than that of protons. This is 
mainly due 
to the difference in the attenuation lengths for protons ($110~Mpc$) and 
photons ($2.6~Mpc$ \cite{PB} and $2.2~Mpc$ \cite{B70}). At higher energy 
the attenuation length for protons dramatically decreases ($13.4~Mpc$ at 
$E=1 \cdot 10^{12}~GeV$) and the fluxes of protons and photons become 
comparable \cite{BV}.

The predictions in Fig.7. are compared with the AGASA data \cite{Nagano}.
The agreement is rather good for $E>2 \cdot 10^{19}~eV$. The contribution 
of UHE photons increases the total flux at $E> 2\cdot 10^{20}~eV$. The 
contribution of low-energy component (e.g. from radiogalaxies \cite{BiSt})
can easily improve the agreement at lower energies.

The radiation from Topological Defects can explain the diffuse gamma-ray 
background above $10~GeV$ \cite{BaShSt}.

{\em Superheavy relic particles} can be sources of UHE CR \cite{KR,BKV,FKN}.
In this scenario Cold Dark Matter (CDM) have a small admixture 
of long-lived superheavy particles. These particles must be heavy,
$m_X > 10^{12}~GeV$, long-lived $\tau_X > t_0$, where $t_0$ is the age 
of the Universe, and weakly interacting. The required life-time 
can be 
provided if this particle has (almost) conserved quantum number broken 
very weakly due to warmhole \cite{BKV} or instanton \cite{KR} effects.
Several mechanisms for production of such particles in the early Universe 
were identified. Like other forms of non-dissipative CDM , X-particles 
must accumulate in the halo of our Galaxy \cite{BKV} and thus they produce 
UHE CR without GZK cutoff and without appreciable anisotropy. 

The more detailed discussion as well as calculated fluxes of UHE protons 
and photons are given in the paper by M. Kachelriess (these Proceedings).

The characteristic and unavoidable feature of this model is an excess 
of gamma-ray flux over the nucleon flux. It follows from the more effective 
production of pions than nucleons in the QCD cascades from the decay of $X$
-particles and from absence of absorption in the Galactic halo.

The spectrum of the observed EAS is formed due to fluxes of gamma-rays and 
nucleons. The gamma-ray contribution to this spectrum is rather complicated.
In contrast to low energies, the photon-induced showers at 
$E>10^{18}$~eV have the low-energy muon component as abundant as that 
for nucleon-induced showers \cite{AK}. The  
shower production by the photons is, in principle, suppressed by the 
LPM effect \cite{LPM} and by absorption in geomagnetic field. However, as 
was noted 
in \cite{AK} cascading in geomagnetic field results in arrival of 
a bunch of photons (each with a smaller energy than the primary one) at 
the top of atmosphere. They produce one shower with LPM effect  
reduced because of the smaller energies of photons in the bunch (see 
\cite{ps} and references therein for further discussion). Experimental 
discrimination of gamma-ray 
induced showers from that produced by protons is very important task.

{\em The UHE carriers} can be, in principle, not only protons and photons, 
but other 
particles, such as neutrinos, gluinos \cite{Farr,MN,BK}, 
neutralinos \cite{BK} and monopoles \cite{Weil,MN}. In the next section 
we shall shortly discuss discuss the Lightest Supersymmetric Particle as a 
candidate 
for a carrier of a signal in UHECR detectors.  

\section{\bf LSP as UHE carrier}

The Lightest Supersymmetric Particle (LSP) can be either stable, if R-parity 
is strictly conserved, or unstable, if R-parity is violated. To be able to 
reach the Earth from most remote regions in the Universe, the LSP must have 
lifetime longer than $\tau_{LSP} \geq t_0/\Gamma$, where $t_0$ is the 
age of the Universe and $\Gamma=E/m_{LSP}$ is the Lorentz-factor of the LSP.
In case $m_{LSP} \sim 100~GeV$, $\tau_{LSP} > 1~yr$.

Theoretically the best motivated candidates for LSP are the neutralino and 
gravitino; the latter is practically undetectable as UHE particle.

In all elaborated SUSY models the gluino is not the LSP. From experimental 
point of view there is some controversy if
the low-mass window $1~GeV \leq m_{\tilde g} \leq 4~GeV$ for
the gluino is still allowed \cite{pdg,aleph}. Recently 
the light-gluino was claimed to be ruled out on the basis of its contribution 
to $\beta$ function for the strong interaction \cite{CF}. Finally, there is 
one more argument about light gluino \cite{VO}:
If gluino is stable or quasi-stable and if the lightest gluino-baryonic 
state is $\tilde{g}uud$ , this heavy hydrogen is overproduced by cosmic rays 
in the Earth atmosphere. 
Nevertheless, we shall discuss the gluino as UHE carrier \cite{Farr,MN,BK}.
We shall refer to any colorless hadron containing gluino as 
$\tilde{g}$-hadron. The lightest $\tilde{g}$-hadron is most probably 
{\em glueballino} $\tilde{g}g$. It is stable if gluino is LSP. 
The lightest gluino-baryonic state, {\em gluebarino}, is almost stable 
because of very weak violation of baryonic number.   
Light glueballinos as 
UHE particles with energy $E \geq 10^{16}$~eV were  considered in 
some detail in the literature in connection with Cyg X-3 \cite{aur,BI}.

UHE LSP are most naturally produced at the decays of unstable superheavy 
particles, either from TD or the relic ones \cite{BK}. 

The QCD parton cascade is not a unique cascade process. A cascade 
multiplication of partons at the decay of superheavy particle appears 
whenever a probability of production of extra parton has the terms 
$\alpha \ln Q^2$ or $\alpha \ln^2 Q^2$, where $Q$ is a maximum of parton 
transverse momentum. Regardless of smallness of 
$\alpha$, the cascade develops as far as  $\alpha \ln Q^2 \geq 1$. Therefore, 
for extremely large $Q^2$ we are interested in, a cascade develops due to
parton multiplication through $SU(2)\times U(1)$ interactions as well (see 
\cite{BK} for calculations). LSP take away a 
considerable fraction of the total energy ($\sim 40\%$).  

{\em Neutralino} as UHE carrier has no much advantages over neutrino: 
the neutralino fluxes, produced at the decay of superheavy particles, are 
less than neutrino fluxes and neutralino-nucleon cross-section at very 
high energy is less than that for neutrino \cite{BK}.

{\em Light gluino} is effective as the  UHE carrier.
The energy losses of 
glueballino on a way from  
a source to the Earth is less than for a proton. The dominant energy loss 
is due to pion production in collisions with microwave 
photons. Pion production effectively starts at the same Lorentz-factor as 
in the case of the proton. This implies that the energy of the GZK 
cutoff is a factor $m_{\tilde{g}}/m_p$ higher than in case of the proton. 
The attenuation length also increases because the fraction of energy lost 
near the threshold is small, $\mu/m_{\tilde{g}}$, where $\mu$ is a pion 
mass. Therefore, even for light glueballino, $m_{\tilde{g}} \geq 
2~GeV$, the steepening of the spectrum is less pronounced than for 
protons \cite{BK}.
 
A very light UHE glueballino interacts with the nucleons in the 
atmosphere similarly to UHE proton. The cross-section for heavy glueballino 
 with $m_{\tilde{g}} > 150~GeV$ is small for large energy transfer needed 
for production of extensive air showers, and thus these particles cannot be 
responsible for the observed UHE events. 

Thus, only UHE gluino from the low-mass window 
$1~GeV \geq m_{\tilde{g}}\leq 4~GeV$ could be a candidate for observed 
UHE particles, but it is disfavored by the arguments given above. 

\section{\bf Conclusions}

At $E\geq 1\cdot 10^{19}~eV$ a new component of cosmic rays with a flat 
spectrum is observed. Two highest energy events have 
$E \approx 2-3\cdot 10^{20}~eV$. According to the Fly's Eye and 
Yakutsk data the chemical composition is better described by protons than 
heavy nuclei. The AGASA data are consistent with isotropy in arrival 
of the particles, though theoretical analysis reveals some correlation 
of arrival direction with Local Supercluster plane. AGASA has observed 
clustering of UHE events: three pairs of particles with small angular 
separation.  

The galactic origin of UHECR is disfavored: the maximal observed energies
are less than that known for the galactic sources, and the strong Galactic disc 
anisotropy is predicted even for the extreme magnetic fields in the disc 
and halo.

The signature of extragalactic UHECR is GZK cutoff. The position of 
steepening is model-dependent value. For the Universe uniformly filled with 
sources, the steepening starts at $E_{bb} \approx 3\cdot 10^{19}~eV$ and 
has $E_{1/2} \approx 6\cdot 10^{19}~eV$ (the energy at which spectrum 
becomes a factor of two lower than a power-law extrapolation from lower 
energies). The spectra of UHE nuclei exhibit steepening approximately at the 
same energy as protons. UHE photons have small absorption length due to 
interaction with radio background radiation. 

The extragalactic astrophysical sources theoretically studied so far, 
have either too small $E_{max}$ or are located too far away. The Local 
Supercluster (LS) model can give spectrum with $E_{1/2} \sim 10^{20}~eV$, if 
density contrast for the sources (the ratio of densities inside LS and 
outside) is larger than 10. 

Topological Defects naturally produce particles with extremely high 
energies, much in excess of what is presently observed. However, the fluxes 
from most known TD are too small. So far only necklaces  and 
monopole-antimonopole pairs  can provide the observed flux of UHE CR. 

Another promising sources of UHE CR are relic superheavy particles. 
 These particles should be clustering in the halo of 
our Galaxy, and thus UHECR produced at their decays do not 
have the GZK cutoff. The signatures of this model are dominance of 
photons in the primary flux and Virgo cluster as a possible discrete source.

Apart from protons and photons, the light gluinos can be successful UHE 
carriers, but they are disfavored in mass interval at interest.

\section{\bf Acknowledgements}

I am grateful to my co-authors Michael Kachelriess and 
Alex Vilenkin for many useful discussions when preparing this talk. Michael 
Kachelriess is thanked also for the help in the calculations which were 
performed for this presentation. I much benefited from conversations with 
J. Cronin and M.Nagano.


\begin{thebibliography}{99}

\bibitem{Hay96} H.Hayashida et al., Proc. of Int. Symposium "Extremely 
High Energy Cosmic Rays" (ed. M.Nagano), Univ. of Tokyo (1996) 17.

\bibitem{GZK} K.Greisen, Phys. Rev. Lett. 16 (1966) 748; G.T.Zatsepin and 
V.A.Kuzmin, JETP Lett. 4 (1966) 78.

\bibitem{BBDGP} V.S.Berezinsky, S.V.Bulanov, V.A.Dogiel, V.L.Ginzburg and 
V.S.Ptuskin, Astrophysics of Cosmic Rays, chapter 4, North-Holland, 1990. 

\bibitem{HS85} C.T.Hill and D.N.Schramm, Phys. Rev. D 31 (1995) 564.

\bibitem{BG87} V.S.Berezinsky and S.I.Grigorieva Sov. Phys. JETP 66 (1987) 
457 and Astron. Astroph. 199 (1988) 1.

\bibitem{B70} V.S.Berezinsky , Sov. J. Nucl. Phys. 11 (1970) 222.

\bibitem{PB} R.J.Protheroe and P.L.Biermann, Astrop. Phys. 6 (1996) 45.

\bibitem{akeno} S.Yoshida et al., Astrop. Phys., {\bf 3}, 105, (1995).

\bibitem{HFE} D.J.Bird et al, Ap.J. 424 (1994) 491.

\bibitem{HAGASA} N.Hayashida et al., Phys. Rev. Lett. 73 (1994) 3491.

\bibitem{HP} G.Canningham et al., Ap.J. 236 (1980) L71.

\bibitem{Syd} M.M.Winn et al. J. Phys. G: Nucl. Phys. 12 (1986) 653.

\bibitem{Lins97} Similar remark was made recently by J. Linsley.

\bibitem{Gl} A.V.Glushkov et al., J. Nucl. Phys. Preprint (1994).

\bibitem{Yak} B.N.Afanasiev et al, Proc. of Tokyo Workshop on Techniques 
for the study of Extremely High Energy Cosmic Rays, (eds G.Loh et al),
Institute for Cosmic Ray Research, Tokyo 1993, 35. 

\bibitem{SW} T.Stanev et al, Phys. Rev. Lett. 75 (1995) 3056.

\bibitem{AGASA1} N.Hayashida et al., Phys. Rev. Lett. 77 (1996) 1000.

\bibitem{SLO} G.Sigl, M.Lemoine and A.Olinto, astro-ph/9704204

\bibitem{FE} D.Bird et al., Phys. Rev. Lett. 71 (1993) 4301.

\bibitem{AGASA2} N.Hayashida et al., J.Phys. G: Nucl. Phys. 21 (1995) 1101.


\bibitem{Yak1} B.N.Afanasiev et al, Proc. of Int. Workshop "Extremely High 
Energy Cosmic Rays" (ed. M.Nagano), Institute for Cosmic Ray Research, Tokyo
(1996) 32.

\bibitem{Ces} C.J.Cesarsky and P.O.Lagage, Astron. Astrophys. 125 (1983) 
294.

\bibitem{BiVo} H.J.Volk and P.L.Biermann, Ap.J.Lett. 333 (1988) 65. 

\bibitem{BC} P.L.Biermann and J.P.Cassinelli, Astron Astroph., 277 (1993) 
691.

\bibitem{JM} J.R.Jokipii and G.Morfill, Ap.J. 312 (1987) 170.


\bibitem{SC} A.G.K.Smith and R.W.Clay, Austr. J.Phys. 43 (1990) 373.

\bibitem{BGR} V.Berezinsky, S.Grigorieva, A.Mikhailov, H.Rubinstein, A.
Ruzmaikin, D.Sokoloff, A.Shukurov, Proc. of Int. Workshop "Astrophys. 
Aspects of Most Energetic Cosmic Rays" (eds N.Nagano and F.Takahara), 
World Scientific, (1991) 134.

\bibitem{GW} M.Giler, J.L.Osborne, J.Wdowczyk, and M.Zielinska, 23 Int.
Cosm. Ray Conf., (Calgary) 2 (1993) 81. 

\bibitem{PP} D.N.Pocherkin, V.S.Ptuskin, S.I.Rogovaya and V.N.Zirakashvili,
24th Int. Cosm. Ray Conf. (Rome) 3 (1995) 136.

\bibitem{SS} D.D.Sokoloff and A.M.Shukurov, Nature 347 (1990) 51.

\bibitem{Hay} S.Hayakawa, Atroph. Sp. Sci 229 (1972) 237.

\bibitem{BP1} V.S.Berezinsky and O.F.Prilutsky, Astroph. Sp. Sci., 21 
(1973) 475.


\bibitem{NMA} C.A.Norman, D.B.Melrose, and A.Achtenberg, Ap. J. 454 (1995) 60.

\bibitem{BiSt} P.L.Biermann and P.A.Strittmatter, Ap.J. 322 (1987) 643.

\bibitem{IA} W.H.Ip and W.I.Axford, Astrophysical Aspects of the Most 
Energetic Cosmic Rays (ed. M.Nagano), World Scientific, 273 (1991).

\bibitem{RB} J.P.Rachen and P.L.Biermann, Astron. Astroph. 272 (1993) 161.

\bibitem{Bla} R.D.Blandford, Mon.Not.Roy.Astr.Soc. 176 (1976) 465.

\bibitem{Lov} R.V.E.Lovelace, Nature 262 (1976) 649.

\bibitem{Fer} E.Fermi, Phys.Rev. 75 (1949) 1169.

\bibitem{BeKi} M.C.Begelman and J.G.Kirk, Ap.J. 353 (1990) 66.

\bibitem{Ho} M.Hoshino, J.Arons, Y.Gallant, and A.Langdon, Ap.J. 390 (1992) 
454.


\bibitem{Vi} M.Vietri, Ap.J. 453 (1995) 883.

\bibitem{Wa} E.Waxman, Phys. Rev. Lett. 75 (1995) 386.

\bibitem{BG} V.S.Berezinsky and S.I.Grigorieva, Proc. 16th Int. Cosm. Ray 
Conf. (Kyoto) 2 (1979) 81.
\bibitem{KK} K.T.Kim, P.P.Kronberg, G.Giovannini, and T.Venturi, Nature 341 
(1989) 720.

\bibitem{Book}
A. Vilenkin and E.P.S. Shellard, Cosmic Strings and Other Topological
Defects, Cambridge University Press, Cambridge, 1994;
M.B. Hindmarsh and T.W.B. Kibble, Rep. Prog. Phys. {\bf 55}, 478 (1995).

\bibitem{HS}
C.T. Hill, D.N. Schramm and T.P. Walker, Phys. Rev. D36 (1987) 1007;

\bibitem{BHSS} P. Bhattacharjee, C.T. Hill and D.N. Schramm, Phys. Rev. Lett. 
69 (1992) 567;
G. Sigl, D.N. Schramm and P. Bhattacharjee, Astropart. Phys. 2 (1994)
401;  

\bibitem{Sigl} G.Sigl, astro-ph/9611190.


\bibitem{BR} P.Bhattacharjee and N.C.Rana, Phys. Lett. {\bf B 246}, 365 
(1990).

\bibitem{Vincent} G.Vincent, N.Antunes and M.Hindmarsh, hep-ph/9708427.

\bibitem{Bran}
R.Brandenberger, Nuclear Physics, {\bf B 293}, 812 (1987).

\bibitem{GK}
A.J. Gill and T.W.B. Kibble, Phys. Rev. D50 (1994) 3660.

\bibitem{BBM}
J.H. MacGibbon and R.H. Brandenberger, Nucl. Phys. {\bf B331}, 153 (1990);
P. Bhattacharjee, Phys. Rev. {\bf D40}, 3968 (1989).

\bibitem{Witten}
E. Witten, Nucl. Phys. {\bf B249}, 557 (1985).

\bibitem{SJSB}
V.Berezinsky and A.Vilenkin, in preparation.

\bibitem{Hill}
C.T. Hill, Nucl. Phys. {\bf B224}, 469 (1983).

\bibitem{BS}
P. Bhattacharjee and G. Sigl, Phys. Rev. {\bf D51}, 4079 (1995).

\bibitem{BMV}
V. Berezinsky, X. Martin and A. Vilenkin, Phys. Rev {\bf D 56}, 2024 (1997)
.
\bibitem{BBV} V.Berezinsky, P.Blasi and A.Vilenkin, in preparation.

\bibitem{BV} V.Berezinsky and A.Vilenkin, Phys. Rev. Lett. 79 (1997) 5202.

\bibitem{Nagano} We thank M.Nagano for providing us with these data.

\bibitem{BaShSt} P.Bhattacharjee, Q.Shaffi and F.W.Stecker, hep-ph/9710533.

\bibitem{KR} V.A.Kuzmin and V.A.Rubakov , Talk at the Workshop 
"Beyond the Desert", Castle Rindberg 1997, astro-ph/9709187.

\bibitem{BKV} V.Berezinsky, M.Kachelriess and A.Vilenkin,
Phys. Rev. Lett. 79 (1997) 4302.

\bibitem{FKN} P.H.Frampton, B.Keszthelyi, and Y.J.Ng, astro-ph/9709080.

\bibitem{AK}
F. A. Aharonian, B. L. Kanevsky and V. A. Sahakian, 
J. Phys. {\bf G17}, 1909 (1991).

\bibitem{LPM}
L. D. Landau and I. Pomeranchuk, Dokl. Akad. Nauk SSSR, {\bf92}, 535 (1953);
A. B. Migdal, Phys. Rev.,{\bf 103}, 1811 (1956).

\bibitem{ps}
R. Protheroe and T. Stanev, Phys. Rev. Lett. {\bf 77}, 3708 (1996) and 
erratum.

\bibitem{Farr} D.J.H. Chung, G.R.Farrar and E.W.Kolb, astro-ph/9707036.

\bibitem{MN} R.N.Mohapatra and S.Nussinov, hep-ph/9708497.

\bibitem{BK} V.Berezinsky and M.Kachelriess, hep-ph/9709485.

\bibitem{Weil} T. Kephart and T.Weiler, Astrop. Phys. {\bf 4}, 271 (1996).






\bibitem{pdg}
Particle Data Group, Phys. Rev. {\bf D54} (1996) 1.

\bibitem{aleph}
Aleph collaboration, CERN-PPE-97/002, to be published in Z. Phys. C;
G. R. Farrar, hep-ph/9707467.

\bibitem{CF} F.Csikor and Z.Fodor, Phys. Rev. Lett. 78 (1997) 4335.

\bibitem{VO}
M. B. Voloshin and L. B. Okun, Sov. J. Nucl. Phys. 43 (1986) 495.


\bibitem{aur}
G. Auriemma, L. Maiani and S. Petrarca, 
Phys. Lett. {\bf B164} (1985) 179.

\bibitem{BI}
V. S. Berezinskii and B. L. Ioffe, 
Sov. Phys. JETP {\bf 63} (1986) 920.


\end{thebibliography}
\end{document}